\documentclass[aps,prd,preprint,superscriptaddress,tightenlines,nofootinbib]{revtex4}

\usepackage{graphicx}
\usepackage{dcolumn}
\usepackage{bm}

\begin{document}


\preprint{CLNS 03/1815}
\preprint{CLEO 03-02}

\title{\large Study of the Charmless Inclusive $B \to \eta'X$ Decay}

\author{G.~Bonvicini}
\author{D.~Cinabro}
\author{M.~Dubrovin}
\author{S.~McGee}
\affiliation{Wayne State University, Detroit, Michigan 48202}
\author{A.~Bornheim}
\author{E.~Lipeles}
\author{S.~P.~Pappas}
\author{A.~Shapiro}
\author{W.~M.~Sun}
\author{A.~J.~Weinstein}
\affiliation{California Institute of Technology, Pasadena, California 91125}
% ...
\collaboration{CLEO Collaboration} 
\noaffiliation

\date{05 March 2003}

\begin{abstract}
Based on a measurement of high momentum $\eta'$ production in $B$
decays, we determine the charmless inclusive $B\to\eta'X_{nc}$
branching fraction in the lab-frame momentum interval
$2.0<P_{\eta'}<2.7$~GeV$/c$. Using $9.7\times10^6 B\bar{B}$ pairs
collected at the $\Upsilon(4S)$ center-of-mass energy with the
CLEO II and II.V detector configurations, we find ${\cal B}(B \to
\eta' X_{nc}) = [4.6 \pm 1.1 \pm 0.4 \pm 0.5]\times 10^{-4}$ in
the $2.0<P_{\eta'}<2.7$~GeV$/c$ momentum range, where the
uncertainties are statistical, systematic, and from subtraction
of background from $B$ decays to charm, respectively.
\end{abstract}
\maketitle

The flavor changing neutral current decays $b \to s g$ are
forbidden at tree level in the Standard Model (SM), and hence at
lowest order can only occur at the loop level. Interest in the
charmless inclusive decay $B\to\eta' X$ (which we denote as
$B\to\eta' X_{nc}$) arises because it is expected to be dominated
by the $b \to s g$ transition followed by fragmentation of the
gluon into $\eta'$ via QCD anomaly coupling~\cite{as}-\cite{ap}
and formation of multiparticle states $X$ by the $s$ quark. The
amplitude of these gluonic penguin decays may receive significant
contributions from diagrams with virtual non-SM particles in the
loop.

CLEO previously reported an unexpectedly large rate for high
momentum $\eta'$ production from $B$ decays~\cite{browder}. That
result was based on $4.7$ fb$^{-1}$ of total luminosity, taken
both on the $\Upsilon$(4S) resonance and at a center-of-mass
energy 60~MeV below the resonance, which is below the $B\bar{B}$
production threshold.  These datasets are referred to as
``on-resonance'' and ``off-resonance'' respectively. CLEO found
an inclusive $B\to\eta' X_{nc}$ branching fraction~\cite{browder}
of $[6.2 \pm 1.6($stat$) \pm
1.3($syst$)^{+0.0}_{-1.5}($bkg$)]\times 10^{-4}$ for
$2.0<P_{\eta'}<2.7$~GeV/$c$.

In this paper we present a new measurement of $\eta'$ production
from $B$ decays in the momentum range
$2.0<P_{\eta'}<2.7$~GeV/$c$.  Then, by subtracting background from
$B$ to charm decays, we use this result to determine $B\to \eta'
X_{nc}$. The result is based on $9.1$ fb$^{-1}$ of on-resonance
data and $4.4$ fb$^{-1}$ of off-resonance data. The analysis
method is improved over CLEO's previous $B\to\eta' X_{nc}$
analysis~\cite{browder}, and now uses a combination of the
``pseudoreconstruction'' and neural-network/shape-variable
approaches that have been used in the CLEO analyses of $b\to
s\gamma$~\cite{ernst},~\cite{cleobsg}. These strategies are used
to isolate the signal and to suppress the contribution from
continuum $\eta'$ production. We first search for
$B\to\eta'X_{nc}$ candidates that are consistent with one of the
$B$ meson multiparticle decays. We then estimate the background
from $B$ decays to charm via Monte Carlo technique.  These new
results include the data used in the previous analysis and the
results presented here supersede that measurement.

The data used for this analysis were collected with the CLEO
detector~\cite{cleo} at the Cornell Electron Storage Ring, a
symmetric $e^+e^-$ collider. The CLEO detector measures charged
particles over $95\%$ of $4\pi$ steradians with a system of
cylindrical drift chambers. (For 2/3 of the data used here, the
innermost tracking chamber was a 3-layer silicon vertex
detector.) Its CsI calorimeter covers $98\%$ of $4\pi$. Charged
particles are identified by specific ionization measurement
($dE/dx$) in the outermost drift chamber and by time-of-flight
counters placed just beyond the tracking volume. Muons are
identified by their ability to penetrate the iron yoke of the
magnet. Electrons are identified by the ratio of their shower
energy to track momentum, by track-cluster matching, and by their
shower shape.

We select events using a standard set of CLEO criteria for
hadronic final states, and then search for candidate $\eta'$
mesons by reconstructing the $\eta' \to \eta \pi^+ \pi^-$, $\eta
\to \gamma \gamma$ mode, where the $\gamma \gamma$ forming the
$\eta$ candidate must have an invariant mass within 30~MeV of the
nominal $\eta$ mass. We then form the mass difference between the
reconstructed $\eta'$ and $\eta$ masses to improve resolution.
The mass difference $M(\pi^+ \pi^-\gamma \gamma)-M(\gamma
\gamma)$ must be within 50~MeV of the nominal mass difference
$\Delta M_{PDG} = M_{\eta'} - M_{\eta} = 410.5$~MeV~\cite{pdg}.
We restrict the $\eta'$ momentum to $P_{\eta'}> 1.6$~GeV/$c$.

Using reconstruction we attempt to identify events in which a $B$
decay produces a strange quark recoiling against an $\eta'$. The
reconstruction is done by forming combinations of a charged kaon
or a $K^0_S \to \pi^+\pi^-$, an $\eta'$ candidate, and $n$ pions
where  $n \le 4$ (at most one of these pions is allowed to be
neutral); a total of eighteen decay modes and their charge
conjugates are considered. For each $B$ candidate we calculate
the momentum $P$, energy $E$, and beam-constrained mass $M \equiv
\sqrt{E_{\it beam}^2 - P^2}$. We then form the $\chi^2_B$ of the
reconstruction:
$$\chi^2_B \equiv \left(\frac{E-E_{\it beam}}{\sigma_E}\right)^2 +  \left(\frac{M-M_B}{\sigma_M}\right)^2,$$
where $\sigma_E = 40$ MeV and $\sigma_M = 4$ MeV.

We only consider candidates with $\chi^2_B < 20$, and if an event
has more than one candidate with $\chi^2_B < 20$ we choose the
candidate with the lowest $\chi^2_B$. This reconstructed X system
may include strange, charm, or light quarks. Events with only
light quarks in the recoil system ($X_{d(u)}$), however, will have
slightly lower efficiency than the signal $X_s$ cases. Events with
charm in the recoil system will have comparable efficiency to the
$X_s$ cases, and we will later subtract these as background. In
cases where all decay products of $X$ are identified correctly
($50\%$ of the time) the $X$ mass ($M(X)$) resolution varies from
20 to 30~MeV for low and high $M(X)$ respectively. The $M(X)$
resolution becomes 200 to 300~MeV if there are missing or extra
particles in the $X$ reconstruction. Given a reconstructed
candidate, we use $\chi^2_B$ and $|$cos $\theta_{tt}|$ as
variables for continuum suppression, where $\theta_{tt}$ is the
angle between the thrust axis of the candidate $B$ and the thrust
axis of the rest of the event.

We also use event shape variables and the presence or absence of
a lepton (electron or muon) to further suppress the continuum
background. Specifically, we use a neural network optimized on
signal and continuum Monte Carlo to combine the following event
shape variables into a single variable $r_{shape}$: the normalized
Fox-Wolfram second moment $R_2$~\cite{fox},
$S_\bot$\footnote{$S_\bot$ is defined as the sum of the magnitudes
of components of momenta perpendicular to the direction of the
$\eta'$ for all particles more than $45^\circ$ from the $\eta'$
axis, divided by the sum of the magnitudes of momenta of all
particles other than the $\eta'$.}, and the energies in
$20^\circ$ and $30^\circ$ cones, parallel and antiparallel to the
$\eta'$ direction.  If the event contains a lepton then we also
use the momentum of the lepton, $P_l$, and the angle between the
lepton and the $\eta'$, $\theta_{l\eta'}$.

We thus have two types of events: those with both a
pseudoreconstruction and a lepton, for which we use $r_{shape}$,
$\chi^2_B$, $|$cos $\theta_{tt}|$, $P_l$, and $\theta_{l\eta'}$;
and events with only a pseudoreconstruction, for which we use
$r_{shape}$, $\chi^2_B$, and $|$cos $\theta_{tt}|$. For each of
these two cases the available variables are combined using a
neural network that has been optimized using signal and continuum
Monte Carlo samples. For each of the two resulting networks, we
can maximize the statistical strength by converting the network
output $r$ into a weight, $w(r)=s(r)/[s(r)+(1+a)b(r)]$, where
$s(r)$ and $b(r)$ are the expected yields for signal and for
continuum background respectively, $r$ is the net output, and $a$
is the luminosity scale factor between on-resonance and
off-resonance data samples ($a\approx 2.0$).

The above choice of weights minimizes the expected statistical
error on the $B\to\eta'X_{nc}$ yield after off-resonance
subtraction.  To reduce the systematic error that arises from
having some efficiency dependence on the reconstructed $M(X)$
value, however, we adjust the weight event-by-event based on the
measured value of $M(X)$. The factor for this adjustment was
obtained from signal Monte Carlo by fitting the measured
efficiency dependence on $M(X)$ to a straight line. By using these
adjusted weights we slightly increase our expected statistical
error, but decrease a systematic error.

We obtain yields from both on- and off-resonance data by fitting the
distributions of the mass difference $M(\pi^+ \pi^-\gamma
\gamma)-M(\gamma \gamma) - \Delta M_{PDG}$. We use Gaussian and
linear functions for the $\eta'$ signal and combinatorial
background respectively.  In these fits, we constrain the mean of
the Gaussian to be zero, and the width to be $4.0$~MeV based on
Monte Carlo simulation.

\begin{figure}
\includegraphics*[width=3.75in]{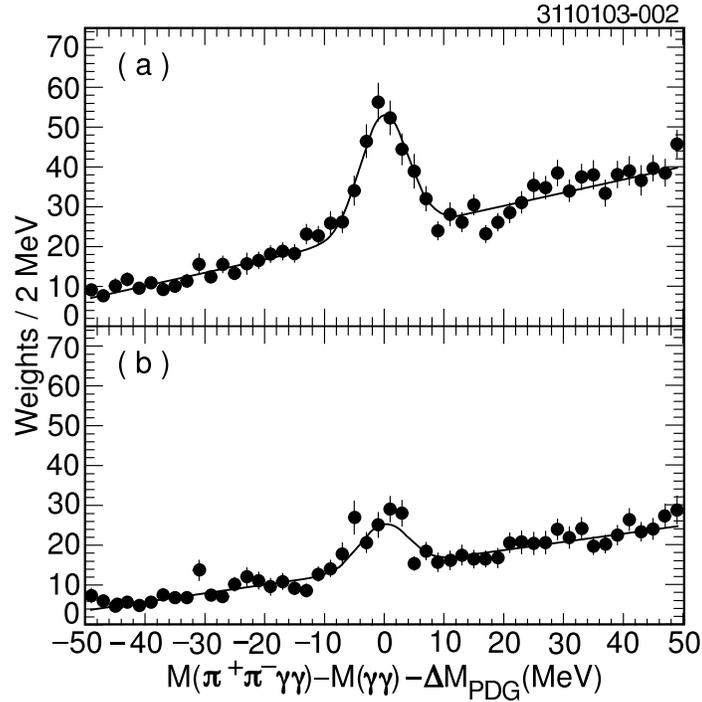}
\caption{
The distribution of $M(\pi^+ \pi^-\gamma \gamma)-M(\gamma
\gamma)-\Delta M_{PDG}$ in the signal region
$2.0<P_{\eta'}<2.7$~GeV/$c$ for (a) on-resonance data and (b)
scaled off-resonance data.
}
\label{fig:dm}
\end{figure}

We scale the off-resonance yield by the on to off ratio of ${\cal
L}/E_{cm}^2$ to account for luminosity and cross-section
differences between the two datasets.  We also scale particle
momenta by the on/off energy ratio to account for the small energy
difference between the on-resonance and off-resonance
data\footnote{The on- and off-resonance datasets have
center-of-mass energies of 10.58 and 10.52~GeV respectively.}. We
then find our $B \to \eta' X$ yield by subtracting this
off-resonance yield from the on-resonance yield. The mass
difference $M(\pi^+ \pi^-\gamma \gamma)-M(\gamma \gamma) - \Delta
M_{PDG}$ in the $\eta'$ momentum range $2.0 < P_{\eta'} <
2.7$~GeV/$c$ is shown in Figure~\ref{fig:dm}, and yields from fits
are tabulated in Table~\ref{tab:yields}. We estimate a systematic
error of $3\%$ from the uncertainty in the fitting procedure.

\begin{figure}
\includegraphics*[width=3.75in]{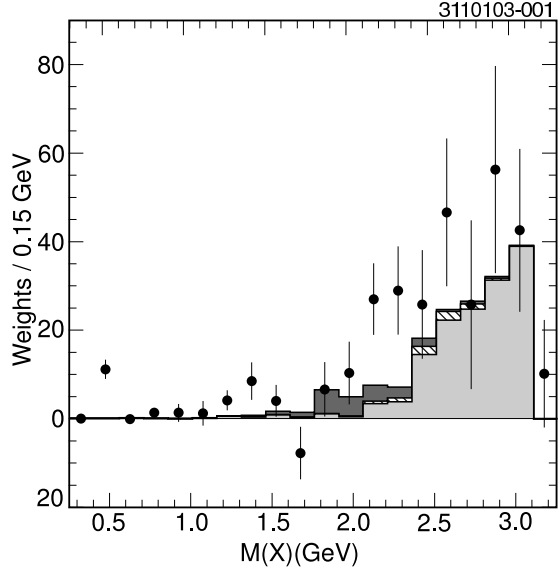}
\caption{The continuum-subtracted and
combinatorial-background-subtracted $M(X)$ distribution (points
with error bars) with estimated background from $B$ decays to
charm (histogram). The various contributions are cascade decays
(light grey) and color suppressed decays $B^0 \to \eta' D^0 +
\eta' D^{0*}$ (dark grey) and $B^0 \to \eta' D^{0**}$ (hatched
area). Ignoring the small smearing effects from the boosted $B$
mesons (300~MeV/c), the mass ranges $M(X)<2.35$~GeV and
$M(X)>2.5$~GeV correspond to momentum ranges in
Table~\ref{tab:yields} of $P_{\eta'}>2.0$~GeV/$c$ and
$P_{\eta'}<1.9$GeV/$c$ respectively.}

\label{fig:mx}
\end{figure}

To search for $B \to \eta' X_{nc}$, we take as signal the momentum
region $2.0 < P_{\eta'} < 2.7$ GeV/$c$ because it is above the
kinematic limit for most $B$ to charm decays.  Some $B$ to charm
decays, however, will still enter this signal region. We estimate
this background by using $B\bar{B}$ Monte Carlo events and by
measuring the data yield in a control region of $1.6 < P_{\eta'}
< 1.9$ GeV/$c$, chosen because here we expect $B \to \eta'
X_{nc}$ to be much smaller than B decays to charm. The estimated
yields from $B$ backgrounds are tabulated in
Table~\ref{tab:yields}. The continuum-subtracted and
combinatorial-background-subtracted $M(X)$ distribution is shown
in Figure~\ref{fig:mx}. Cascade decays (e.g., $B^0 \to D_{(s)}X,
D_{(s)}\to \eta' Y $) and color allowed decays ($b \to c W, W \to
\eta' + (n\pi)$) have been simulated in the CLEO $B\bar{B}$ Monte
Carlo. We do not use this Monte Carlo prediction directly;
instead, we rescale the CLEO $B\bar{B}$ Monte Carlo yield in the
signal region by the factor of $1.06\pm0.22$ to account for half
of the difference between data and Monte Carlo in the control
region. We include the statistical and systematic uncertainties
from the control region in the error on the scale factor. We
estimate the yield of color suppressed decays with a Monte Carlo
sample that assumed the following branching fractions: ${\cal
B}(B^0 \to \eta' D^0 + \eta' D^{0*}) = (3.1\pm0.9)\times
10^{-4}$~\cite{eckhard} and ${\cal B}(B^0 \to \eta' D^{0**}) =
1/2\ {\cal B}(B^0 \to \eta' D^0 + \eta' D^{0*})= 1.55\times
10^{-4}$. We take the error on ${\cal B}(B^0 \to \eta' D^{0**})$
to be $50\%$ of itself. We use the modified ISGW
model~\cite{Isgur:gb} for the mass distribution of $D^{0**}$.

We determine the efficiency --- defined as weights per $B \to
\eta' X_{nc}$ event generated in the momentum range $2.0 <
P_{\eta'} < 2.7$ GeV/$c$ --- via Monte Carlo simulation. The
hadronization of the quark pair into $X_s$ or $X_{d(u)}$ is done
by the JETSET Monte Carlo. We use a uniform $X_{nc}$ mass
distribution for the multiparticle final states, and assume that
the branching fraction for $B\to\eta' K$ is one eighth of the
total branching fraction $B\to\eta'X_s$ in this momentum
range~\cite{gritsan},\cite{browder}. The detection efficiency is
averaged over charged and neutral $B$ mesons and corrects for
unobserved modes with neutral kaons ($K^0_L$, $K^0_S \to
\pi^0\pi^0$), for modes with a charged kaon and more than one
$\pi^0$ in the final state, and for final states with baryons; it
also includes the product of branching fractions ${\cal
B}(\eta'\to\eta\pi^+\pi^-)\times {\cal
B}(\eta\to\gamma\gamma)=16.96\%$.

To estimate the uncertainty in efficiency due to the choice of a
uniform shape for the $M(X_{nc})$ distribution, we vary the shape
of the $M(X_{nc})$ distribution for the multiparticle final
states from uniform to linear with intercept at the $K-\pi$
threshold, keeping the fraction of $B\to\eta' K$ constant; this
leads to a systematic error of $6.3\%$. The systematic error on
the efficiency includes 3.3\% uncertainty from the event modeling
of $B \to \eta' X_{nc}$ which includes uncertainties in event
shape, hadronization, and other-$B$ modeling, and 2.1\%
uncertainty from the detector performance which includes
uncertainties in the detection efficiencies and resolutions for
tracks and photons.

To obtain the $B \to \eta' X_{nc}$ branching fraction, we take the
background-subtracted yield in the momentum range $2.0 < P_{\eta'}
< 2.7$~GeV/$c$ of $61.2\pm13.9\pm6.6$ weights, where the first
error is statistical and the second is from uncertainty in the
subtraction of the $B$ to charm decay background. The efficiency is
$(6.81\pm0.56)\times10^{-3}$ weights per event, where the error
results from dependence of efficiency on the $M(X_{nc})$
distribution, from $X_{nc}$ hadronization, from other-$B$
modeling, and from detector performance. Our sample contains $9.7$
million $B\bar{B}$ pairs ($\pm 2\%$). We obtain ${\cal B}(B \to
\eta' X_{nc}) = [4.6 \pm 1.1($stat$) \pm 0.4($syst$) \pm
0.5($bkg$)]\times 10^{-4}$ for $2.0 < P_{\eta'} < 2.7$~GeV/$c$.

Also included in our measured branching fraction are components
from $b \to d g$ gluonic penguin decays and $b \to u$ tree level
decays. Within the SM, the contribution from these decays is
expected to be of the order of a few per cent of the $B\to\eta'
X_{nc}$ rate. We determine that the ratio of efficiencies for the
final states that consist only of $d$ and $u$ quarks to the
final states that contain $s$ quark is 0.79. Thus, the branching
fraction we have measured is a weighted sum of branching
fractions ${\cal B}(B \to \eta' X_{nc}) = {\cal B}(B \to \eta'
X_s) + 0.79\ {\cal B}(B \to \eta' X_{d(u)})$.

In summary, we have updated the measurement of the charmless
inclusive decay $B \to \eta' X_{nc}$ for $2.0 < P_{\eta'} <
2.7$~GeV/$c$ using the CLEO II and II.V data sets which contain
$9.7$ million $B\bar{B}$ pairs.  This result is in agreement with
CLEO's previous inclusive measurement and hence confirms its
unexpectedly large rate.  CLEO has also previously
measured~\cite{gritsan} the exclusive decay $B \to \eta' K$ using
the same dataset as the results we report here; that result is
therefore not statistically independent of our new results.

The large $B \to \eta' X_{nc}$ rate can be understood if the
$b\to s g$ rate is larger than expected in the Standard
Model~\cite{kp}. A number of alternative explanations, however,
have also been proposed. One of these is the QCD anomaly
mechanism $b\to s g$, $ g\to\eta' g$ with the $\eta' gg$
form-factor being a slowly-falling function of the gluon
$Q^2$~\cite{as}. A slowly-falling $\eta' g g$ form-factor,
however, is strongly disfavored by the recent CLEO measurement of
the $\Upsilon(1S)\to\eta' X$ spectrum~\cite{stone}; a
rapidly-falling form-factor needed for consistency with the
$\Upsilon(1S)\to\eta' X$ result predicts a $B\to\eta'X$ rate of
$3\times 10^{-5}$~\cite{kagan_beyond_sm}, which is an order of
magnitude smaller than observed. Another possible explanation of
the high rate for $B\to\eta' X_{nc}$ is that the $\eta'$ has a
substantial intrinsic charm content~\cite{hz}. In this case the
$\eta'$ can be produced by the axial vector part of the $b\to
(c\bar{c})s$ process. This explanation is also disfavored by CLEO
measurements. First, the $B\to\eta_c K$ rate is not enhanced
relative to the $B\to J/\psi K$ rate~\cite{cleo_etack}.  Further,
this explanation requires that the $B\to\eta'K^*$ rate be roughly
half the $B\to\eta'K$ rate~\cite{ht},\cite{hz}, which is
inconsistent with
observations~\cite{gritsan},\cite{etak_excl_babar}.

We gratefully acknowledge the effort of the CESR staff in
providing us with excellent luminosity and running conditions. M.
Selen thanks the Research Corporation, and A.H. Mahmood thanks
the Texas Advanced Research Program. This work was supported by
the National Science Foundation, and the U.S. Department of
Energy.

\clearpage

\begin{table}
\begin{center}
\begin{tabular}{|l|c|c|}
\hline
$\eta'$ Momentum Range (GeV/$c$) &    1.6 - 1.9                  &    2.0 - 2.7        \\ \hline
ON                               &    $251.8 \pm 25.9$           &    $149.4 \pm 11.4$ \\
$a\times$OFF                     &    $90.8  \pm 15.3$           &    $55.3  \pm 7.9$ \\ \hline
ON$-a\times$OFF                  &    $161.0 \pm 30.0$           &    $94.1  \pm 13.9$ \\ \hline
Color suppressed decays          &                               &                     \\
$B^0 \to \eta' D^0 + \eta' D^{0*}$
                                 &    $0.0   \pm 0.1$            &    $17.4  \pm 5.0$ \\
$B^0 \to \eta' D^{0**}$          &    $2.8   \pm 0.2$            &    $1.4   \pm 0.7$ \\ \hline
CLEO $B\bar{B}$ Monte Carlo      &    $140.6 \pm 11.0$           &    $13.3  \pm 2.9$  \\
Scaled by control region         &                               &    $14.1  \pm 4.3$ \\ \hline
Sum of $B$ backgrounds             &    $143.4 \pm 11.0$           &    $32.9  \pm 6.6$ \\ \hline
ON$-a\times$OFF$-B$ backgrounds  &    $17.6  \pm 30.0(stat)$     &    $61.2  \pm 13.9(stat)$ \\\hline
\end{tabular}
\caption{Yields(weights) from the fit in the signal
($2.0<P_{\eta'}<2.7$~GeV/$c$) and control
($1.6<P_{\eta'}<1.9$~GeV/$c$) regions. Given are yields
on-resonance, scaled off-resonance, on minus scaled off,
estimated background from color suppressed $B$ decays, estimated
background from cascade and color allowed $B$ decays, sum of $B$
backgrounds, and on minus scaled off minus $B$ backgrounds. The
error on the continuum-subtracted and $B$-backgrounds-subtracted
yield is statistical only and does not include the error from the
background subtraction.} \label{tab:yields}
\end{center}
\end{table}

\clearpage



\end{document}